\documentclass[aps,prl,twocolumn,superscriptaddress,showpacs,amsmath,amssymb,nofootinbib]{revtex4-2}

\usepackage{graphicx}  
\usepackage{dcolumn}   
\usepackage{bm}        
\usepackage{hyperref}  
\usepackage{xcolor}
\usepackage{braket}    

\newcommand{\prlsection}[1]{\textit{#1}---}

\begin{document}


\title{Phase Frustration Induced Intrinsic Bose Glass in the Kitaev-Bose-Hubbard Model}

\author{Yi-fan Zhu}
\affiliation{Department of Physics, Beijing Normal University, Beijing 100875, China}

\author{Shi-jie Yang}
\email{yangshijie@tsinghua.org.cn}
\affiliation{Department of Physics, Beijing Normal University, Beijing 100875, China}


\begin{abstract}
We report an intrinsic ``Bubble Phase'' in the two-dimensional Kitaev-Bose-Hubbard model, driven purely by phase frustration between complex hopping and anisotropic pairing. By combining Inhomogeneous Gutzwiller Mean-Field Theory with a Bogoliubov-de Gennes stability analysis augmented by a novel Energy Penalty Method, we demonstrate that this phase spontaneously fragments into coherent islands, exhibiting the hallmark Bose glass signature of finite compressibility without global superfluidity. Notably, we propose a unified framework linking disorder-driven localization to deterministic phase frustration, identifying the Bubble Phase as a pristine, disorder-free archetype of the Bose glass. Our results provide a theoretical blueprint for realizing glassy dynamics in clean quantum simulators.
\end{abstract}

\maketitle


\prlsection{Introduction}Localization phenomena in quantum many-body systems remain a central theme in modern physics. Since the seminal proposal of Anderson localization \cite{anderson_absence_1958}, disorder has been identified as the key mechanism leading to wavefunction localization and insulating behavior. In the specific context of interacting bosons subject to such random potentials, Fisher \textit{et al.} \cite{fisher_boson_1989} established the concept of the ``Bose Glass'' (BG) as a gapless, compressible, yet insulating Griffiths phase. This phase has been experimentally realized in optical lattices using speckle potentials \cite{fallani_ultracold_2007}. While the canonical BG is typically driven by diagonal disorder (random chemical potentials), subsequent studies have shown that off-diagonal disorder (random hopping amplitudes) can also induce similar glassy states \cite{sengupta_quantum_2007, piekarska_quantum_2018, piekarska_emergence_2022}, reinforcing the paradigm that extrinsic randomness is the prerequisite for glassiness.

However, recent research interests have shifted toward the frontier of disorder-free localization \cite{smith_disorder-free_2017, karpov_disorder-free_2021}. A pivotal question is whether intrinsic mechanisms or lattice geometry alone can generate glass-like behavior or spatial fragmentation in clean systems. Significant progress has been made in systems governed by quasi-periodic potentials, exemplified by the celebrated Aubry-Andr\'e model \cite{aubry_analyticity_1980, roscilde_bosons_2008} and incommensurate moir\'e superlattices \cite{ding_interaction-induced_2025}, where deterministic incommensuration induces localization. Beyond quasi-periodicity, recent studies on commensurate twisted bilayers have revealed that geometric interference alone can drive the system into spatially modulated phases, featuring insulating islands embedded in a superfluid sea \cite{zhang_dipolar_2025}. Furthermore, it has been demonstrated that strictly clean quasicrystals can host intrinsic Bose glass phases driven purely by geometric complexity \cite{ciardi_quasicrystalline_2023}.

Among possible intrinsic mechanisms, phase frustration induced by artificial gauge fields has been shown to generate rich quantum phases \cite{powell_bogoliubov_2011, sachdeva_density_2012}. While lattice geometry or interaction constraints typically lead to exotic ordered states---such as vortex lattices \cite{powell_bogoliubov_2011}, supersolids \cite{sachdeva_density_2012}, or chiral Mott insulators \cite{dhar_bose-hubbard_2012}---they can also fundamentally disrupt global phase coherence. Such phase incompatibility may fundamentally disrupt global coherence, potentially favoring spontaneous spatial fragmentation and giving rise to a glassy state in a fully intrinsic and deterministic setting.

Recently, bosonic analogues of the Kitaev model---featuring both complex hopping and explicit pairing~\cite{zhou_condensates_2011}---have garnered significant attention, particularly following their experimental realizations in one-dimensional optomechanical \cite{slim_optomechanical_2024} and superconducting circuit platforms \cite{busnaina_quantum_2024}. While these studies have revealed rich physics such as chiral transport and Majorana-like correlations \cite{mcdonald_phase-dependent_2018, vishveshwara_z2_2021, wang_quantum_2022, peotta_xyz_2014}, they have been predominantly confined to chains or ladders. In this Letter, we generalize this framework to two dimensions, where the orbital magnetic flux emerges as a decisive degree of freedom. Taking the rational flux $\alpha=1/16$ as a representative example,and supported by a dynamical stability analysis, we report that the intrinsic phase frustration arising from the interplay between the Peierls phases and the anisotropic Kitaev pairing drives the system into a robust ``Bubble Phase.'' This phase displays macroscopic phenomenology strongly reminiscent of a Bose glass: it is compressible ($\kappa > 0$) and globally insulating, with transport blocked by spontaneous spatial fragmentation.


\prlsection{Model}We consider interacting bosons on a two-dimensional square lattice driven by artificial gauge fields. The system is described by the Kitaev-Bose-Hubbard Hamiltonian:
\begin{align}
\hat{H} = &-\sum_{\bm{r}} \sum_{\nu=x,y} \left( J e^{i\theta_{\bm{r},\bm{r}+\hat{\nu}}}\hat{b}_{\bm{r}}^{\dagger}\hat{b}_{\bm{r}+\hat{\nu}} + \Delta e^{i\phi_{\bm{r},\bm{r}+\hat{\nu}}}\hat{b}_{\bm{r}}^{\dagger}\hat{b}_{\bm{r}+\hat{\nu}}^{\dagger} + \mathrm{H.c.} \right) \nonumber \\
&+ \frac{U}{2}\sum_{\bm{r}}\hat{n}_{\bm{r}}(\hat{n}_{\bm{r}}-1) - \mu\sum_{\bm{r}}\hat{n}_{\bm{r}},
\label{eq:hamiltonian}
\end{align}
where $\hat{b}_{\bm{r}}$ ($\hat{b}_{\bm{r}}^{\dagger}$) are bosonic annihilation (creation) operators at site $\bm{r}=(x,y)$, and $\hat{n}_{\bm{r}}$ is the number operator. The system is characterized by the hopping amplitude $J$ and the pairing amplitude $\Delta$, dressed by the Peierls phases $\theta$ and $\phi$, respectively. We adopt the symmetric gauge $\mathbf{A} = \pi \alpha (-y, x, 0)$ to define the magnetic flux $\alpha$ per plaquette. The explicit Peierls phases for the bonds connecting $\bm{r}$ and $\bm{r}+\hat{\nu}$ are given by:
\begin{equation}
\theta_{\bm{r},\bm{r}+\hat{x}} = -\pi \alpha y, \quad \theta_{\bm{r},\bm{r}+\hat{y}} = \pi \alpha x.
\label{eq:phases}
\end{equation}
Specifically, we set the pairing phases to strictly follow the gauge field, i.e., $\phi_{\bm{r},\bm{r}+\hat{\nu}} = \theta_{\bm{r},\bm{r}+\hat{\nu}}$. This specific choice generates intrinsic phase frustration due to the distinct transformation properties of the kinetic and pairing terms under spatial inversion. For a generic bond along the direction $\hat{\nu}$, the kinetic term is directional: a particle hopping forward ($\bm{r} \to \bm{r}+\hat{\nu}$) picks up a phase $+\theta$, whereas the reverse process ($\bm{r}+\hat{\nu} \to \bm{r}$) picks up the conjugate phase $-\theta$. In contrast, the pairing term is bond-symmetric: the creation of a pair on the bond $(\bm{r}, \bm{r}+\hat{\nu})$ involves the same phase factor structure regardless of the bond orientation definition. This fundamental mismatch prevents the system from simultaneously minimizing the kinetic and pairing energies, driving the formation of the intrinsic Bubble Phase.


\prlsection{Phase Diagram}To explore the ground state properties of this two-dimensional frustrated system, we employ the Inhomogeneous Gutzwiller Mean-Field Theory (IGMFT) \cite{buonsante_gutzwiller_2009, suthar_supersolid_2020}. The many-body wavefunction is approximated by a product of local states, $|\Psi_{G}\rangle = \prod_{\mathbf{r}} (\sum_{n=0}^{n_{\text{max}}} f_{n,\mathbf{r}}|n\rangle_{\mathbf{r}})$, where $f_{n,\mathbf{r}}$ are the complex variational parameters satisfying the normalization condition $\sum_n |f_{n,\mathbf{r}}|^2 = 1$. The ground state is obtained by minimizing the energy functional $\langle \Psi_G | \hat{H} | \Psi_G \rangle$ via imaginary-time evolution. While mean-field approaches neglect off-site quantum fluctuations, the choice of IGMFT is dictated by the specific computational challenges imposed by the Hamiltonian. Standard exact techniques face severe bottlenecks: the complex Peierls phases introduce a prohibitive sign problem for Quantum Monte Carlo (QMC), while the two-dimensional geometry imposes steep scaling costs for Density Matrix Renormalization Group (DMRG). IGMFT circumvents these limitations, providing a robust framework that naturally incorporates complex phases and explicitly resolves the spatially inhomogeneous order parameters essential for characterizing the Bubble Phase.

\begin{figure}[t]
    \centering
    \includegraphics[width=\linewidth]{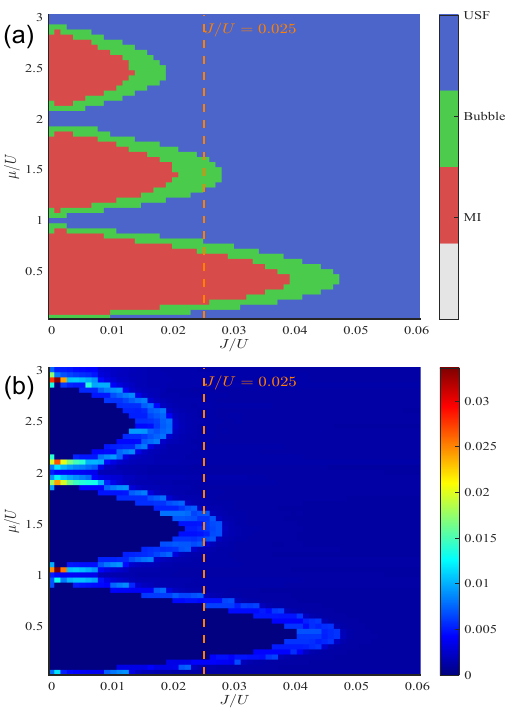}
    \caption{Ground-state phase diagram and coherence signatures for flux $\alpha=1/16$ and pairing amplitude $\Delta/U=0.005$.
    (a) Phase diagram in the $(J/U, \mu/U)$ plane determined by the real-space connectivity of the superfluid order parameter. Three distinct phases are identified: the Mott Insulator (MI), the Unstable Superfluid (USF), and the intervening Bubble Phase. (b) Corresponding momentum-space coherence map quantified by the Inverse Participation Ratio (IPR), $\mathcal{I}_k$. The Bubble phase retains structural coherence (higher IPR) arising from its Bragg peaks, whereas the deep USF regime exhibits a suppressed IPR due to the proliferation of diffuse background modes.}
    \label{fig:phasediagram}
\end{figure}

The resulting ground state phase diagram for a representative flux $\alpha=1/16$ is presented in Fig.~\ref{fig:phasediagram}. To rigorously distinguish the phases, we employ two complementary order parameters focusing on real-space connectivity and momentum-space coherence.

To probe the real-space connectivity of the superfluid regions, we utilize the Hoshen-Kopelman (HK) algorithm \cite{hoshen_percolation_1976}. This technique performs a cluster analysis on the local order parameter field $\psi_{\bm{r}} = \langle \hat{b}_{\bm{r}} \rangle$. Defining two sites as ``connected'' if their correlation $|\psi_{\bm{r}} \psi_{\bm{r}'}|$ exceeds a threshold, the HK algorithm reveals three regimes [Fig.~\ref{fig:phasediagram}(a)]: (i) The Mott Insulator (MI), where no clusters form; (ii) The Superfluid (SF), characterized by a single percolating cluster; and (iii) The Bubble Phase, an intermediate regime where finite-sized, disconnected superfluid clusters emerge but fail to percolate globally.

Complementarily, to distinguish the Bubble phase from the Unstable Superfluid via momentum-space coherence, we analyze the spectral structure of the order parameter using a modified spectral IPR calculated directly from the order parameter's Fourier amplitude $|\tilde{\psi}_{\bm{k}}|$:
\begin{equation}
\mathcal{I}_{\bm{k}} = \frac{\sum_{\bm{k}} |\tilde{\psi}_{\bm{k}}|^2}{(\sum_{\bm{k}} |\tilde{\psi}_{\bm{k}}|)^2}.
\label{eq:IPR_amp}
\end{equation}
Compared to the density-based IPR ($\sim \sum n_{\bm{k}}^2$), this amplitude-based measure provides superior sensitivity to the emerging coherence peaks against the diffuse background characteristic of the frustrated phases, yielding a sharper delineation of the phase boundaries in Fig.~\ref{fig:phasediagram}(b). The resulting map in Fig.~\ref{fig:phasediagram}(b) reveals a counter-intuitive feature: the deep Superfluid regime exhibits a lower $\mathcal{I}_{\bm{k}}$ compared to the Bubble Phase boundary. Detailed inspection shows that while the high-hopping phase retains a condensate peak, it is accompanied by a proliferation of diffuse background modes (precursors to dynamical instability), which suppresses the normalized ratio $\mathcal{I}_{\bm{k}}$. In contrast, the Bubble Phase maintains a cleaner spectral signature within its localized clusters.

Combining these signatures, we confirm that the Bubble Phase resides precisely at the boundary between the MI and the USF phases. This phase structure---where a glassy state intervenes between the insulator and the superfluid---strikingly mirrors the ``Bose glass'' phase observed in models with explicit off-diagonal disorder \cite{sengupta_quantum_2007}, suggesting that phase frustration acts as an intrinsic counterpart to random hopping.

\prlsection{Dynamical Stability and Spectrum}To verify the stability of the Bubble Phase, we investigate the collective excitation spectrum using the quantum Gutzwiller approach developed by Caleffi \textit{et al.} \cite{caleffi_quantum_2020}. Standard BdG approaches rely on expanding boson operators around a condensate mean-field, implicitly assuming that the ground state satisfies the Gross-Pitaevskii equation. However, this assumption fails for the Gutzwiller wavefunction $|\Psi_G\rangle$, which possesses a complex local Fock space structure. To resolve this, we adopt the slave-particle formalism where the Gutzwiller variational parameters are promoted to quantum operators $\hat{f}_{n,\bm{r}}$ satisfying bosonic commutation relations. The physical boson annihilation operator is reconstructed within the Gutzwiller basis as $\hat{b}_{\bm{r}} = \sum_{n} \sqrt{n} \hat{f}_{n-1,\bm{r}}^{\dagger} \hat{f}_{n,\bm{r}}$. Linearizing the Heisenberg equations of motion leads to the generalized BdG eigenvalue problem:
\begin{equation}
\mathcal{M}_{\text{BdG}} \bm{V}_{\nu} = \omega_{\nu} \bm{V}_{\nu},
\label{eq:BdG_matrix}
\end{equation}
where $\mathcal{M}_{\text{BdG}}$ possesses the characteristic block structure enforced by Bose statistics (see Sec.A of Supplemental Material for explicit matrix elements).

A critical challenge in this formalism is the significant enlargement of the configuration space, introducing redundant, non-physical modes that are exact zero-energy eigenstates \cite{caleffi_quantum_2020}. These spurious modes mix with physical low-energy excitations during numerical diagonalization. To decouple the physical sector, we implement the Energy Penalty Method by constructing the effective dynamical matrix $\mathcal{M}_{eff} = \mathcal{M}_{BdG} + \Lambda \mathcal{P}_{spur}$. Here, $\mathcal{P}_{spur}$ is the projection operator onto the spurious subspace, and $\Lambda \gg J$ is a large penalty parameter that shifts them to high energies.A perturbative justification for this energy penalty method, proving the decoupling of the spurious subspace, is provided in Sec. B of the Supplemental Material.

\begin{figure}[t]
    \centering
    \includegraphics[width=\linewidth]{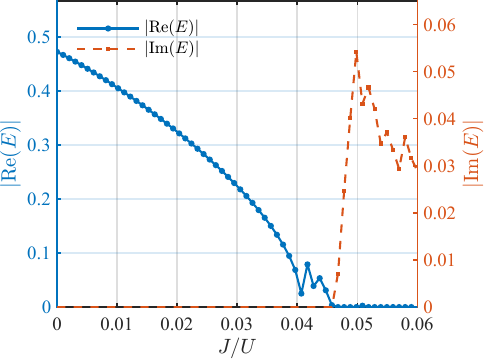}
    \caption{Dynamical stability and excitation spectrum.
    Evolution of the lowest Bogoliubov-de Gennes (BdG) excitation energy as a function of hopping strength $J/U$ (at fixed $\mu/U=0.5$). The solid blue line represents the real energy gap $|\mathrm{Re}(E)|$, while the dashed red line indicates the imaginary part $|\mathrm{Im}(E)|$ associated with dynamical instability. The Bubble phase (for $J/U \lesssim 0.046$) is characterized by a purely real spectrum ($|\mathrm{Im}(E)|=0$) with a finite gap, confirming it is a stable eigenstate.}
    \label{fig:stability}
\end{figure}

The calculated excitation spectrum along a cut at fixed $\mu/U=0.5$ is shown in Fig.~\ref{fig:stability}. This scan traverses three distinct regimes. First, in the Mott Insulator ($J/U \lesssim 0.04$), the lowest excitation energy (blue solid line) is real and finite, decreasing linearly with increasing hopping \cite{fisher_boson_1989}. Second, for the Bubble Phase ($0.04 \lesssim J/U \lesssim 0.046$), the spectrum remains strictly purely real ($|\mathrm{Im}(E)|=0$, red dashed line), providing definitive proof that the Bubble Phase is a dynamically stable eigenstate. Significantly, the system maintains a finite excitation gap. We emphasize that this gap corresponds to the collective quasiparticle excitations at a fixed chemical potential, which are discretized due to the finite-size confinement of the superfluid clusters. Unlike the charge gap in the Mott insulator, this dynamical gap does not imply ground-state incompressibility. Finally, in the Unstable Superfluid regime ($J/U \gtrsim 0.046$), complex eigenvalues emerge ($|\mathrm{Im}(E)| > 0$) upon further increasing hopping, explicitly signaling dynamical instability and the breakdown of the stable Gutzwiller ground state.

These spectral signatures explicitly validate the phase diagram in Fig.~\ref{fig:phasediagram} and rigorously establish the Bubble Phase as a stable and gapped eigenstate of the frustrated Hamiltonian.

\begin{figure}[t]
    \centering
    \includegraphics[width=\linewidth]{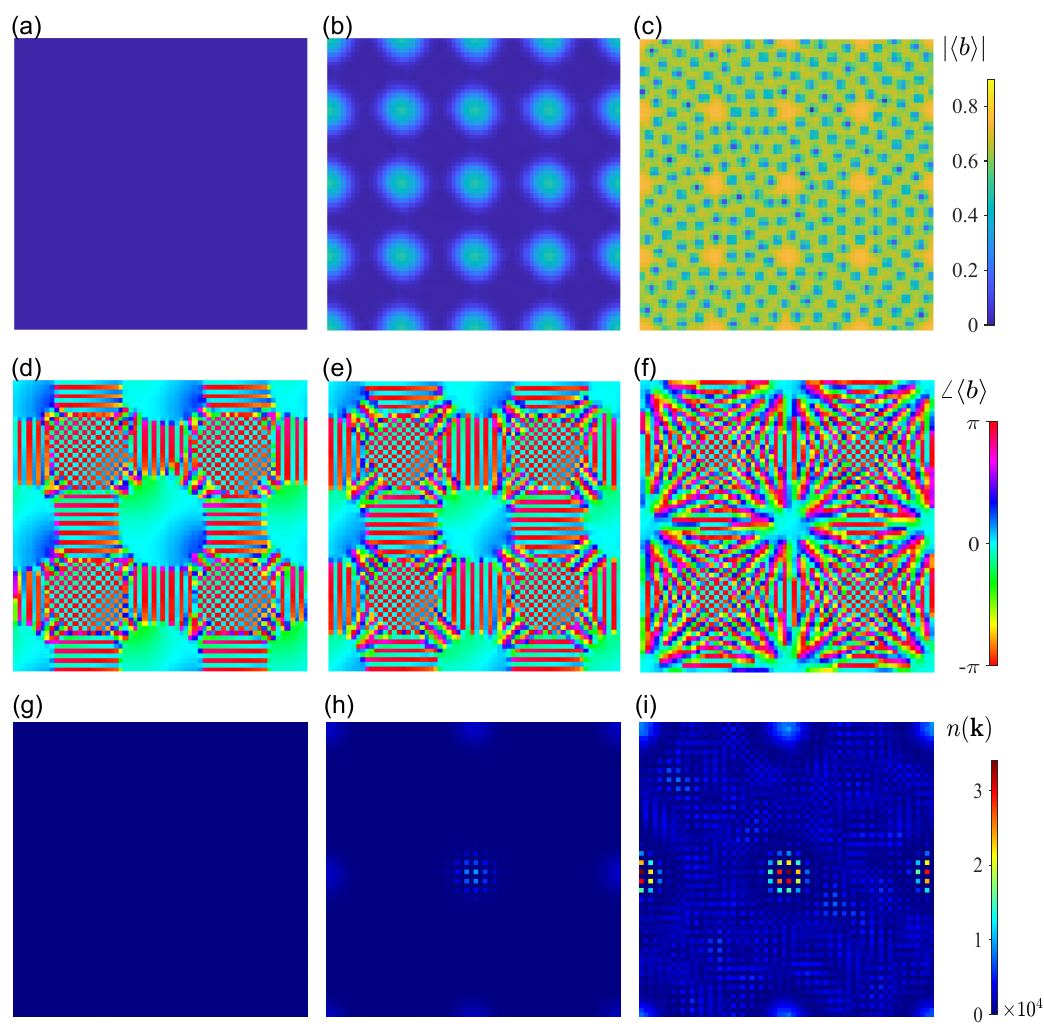}   
    \caption{Microscopic signatures of spontaneous fragmentation.
    Real-space distribution of the local superfluid order parameter amplitude $|\langle b_i \rangle|$ (top row), phase $\arg(\langle b_i \rangle)$ (middle row), and momentum space distribution $n(\mathbf{k})$ (bottom row) for three representative phases on a $64 \times 64$ lattice. (a, d, g) Mott Insulator ($J/U=0.03$): Uniform vanishing amplitude. (b, e, h) Bubble Phase ($J/U=0.045$): Spontaneous fragmentation into localized superfluid ``islands''. Within each bubble, the phase is ordered, but global coherence is disrupted by the insulating domain walls. Momentum distribution (h) shows distinct Bragg peaks but with reduced intensity. (c, f, i) Unstable Superfluid ($J/U=0.06$): Global connectivity with chaotic patterns.}
    \label{fig:microscopic}
\end{figure}

\begin{figure}[t]
    \centering
    \includegraphics[width=\linewidth]{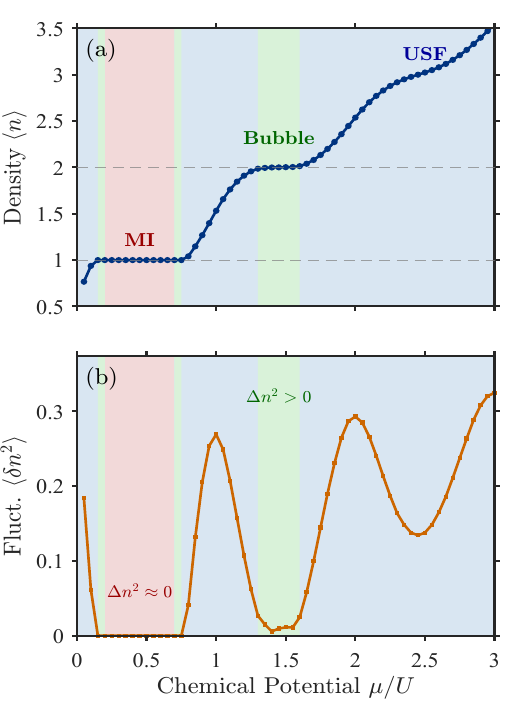}
   \caption{Macroscopic bulk properties at fixed hopping $J/U=0.025$.
    (a) Equation of state showing the average particle number $\langle n \rangle$ vs chemical potential $\mu/U$. The Mott Insulator (red shaded) shows an incompressible plateau ($\kappa \approx 0$). The Bubble Phase (green shaded) exhibits a finite slope, indicating finite compressibility ($\kappa > 0$). (b) Particle number fluctuations $\langle \delta \hat{n}^2 \rangle$. The Bubble Phase supports significant number fluctuations, distinguishing it from the particle-number-squeezed Mott state.}
    \label{fig:eos}
\end{figure}

\prlsection{Bubble Phase Characteristics}Finally, we elucidate the physical characteristics of the Bubble Phase by combining its microscopic texture (Fig.~\ref{fig:microscopic}) with its macroscopic bulk properties (Fig.~\ref{fig:eos}).

The microscopic origin of the insulating behavior lies in the real-space structure of the order parameter $\psi_{\mathbf{r}} = \langle \hat{b}_{\mathbf{r}} \rangle$, shown in Fig.~\ref{fig:microscopic}. Unlike the uniform Mott insulator [Fig.~\ref{fig:microscopic}(a)] or the globally connected Superfluid, the Bubble Phase [Fig.~\ref{fig:microscopic}(b)] spontaneously fragments into an array of localized superfluid islands separated by Mott-insulating domain walls. This fragmentation arises directly from the competition between incompatible local phase configurations. A close inspection of the phase texture in Fig.~\ref{fig:microscopic}(e) reveals the coexistence of distinct local patterns. The Mott insulating regions (where $|\psi_{\mathbf{r}}| \to 0$) are located precisely at the interfaces between these distinct phase domains, acting as barriers to resolve the conflicts between mismatched phase configurations.

This picture of local coherence without global order is corroborated by the momentum-space distribution $n(\mathbf{k})$, displayed in the bottom row of Fig.~\ref{fig:microscopic}. While the Mott phase [Fig.~\ref{fig:microscopic}(g)] is featureless, the Bubble Phase [Fig.~\ref{fig:microscopic}(h)] exhibits distinct Bragg peaks. However, under the same intensity scale, the peak intensity in the Bubble Phase is significantly weaker than that of the Unstable Superfluid [Fig.~\ref{fig:microscopic}(i)]. This suppression of the condensate peak signifies the absence of true long-range off-diagonal order (ODLRO), distinguishing the state from a supersolid and confirming its glassy nature.

The stability of the Bubble Phase relies on avoiding global percolation, which is governed by the interplay between bubble density and bubble size. First, the number of bubbles on an $L \times L$ lattice is fixed at $N_b \approx (L \alpha/\pi)^2$. Consequently, a larger magnetic flux crowds the system, reducing the mean separation between bubbles. Second, the spatial extent of individual bubbles grows with increasing kinetic energy: enhancing either the hopping $J$ or the pairing amplitude $\Delta$ expands the superfluid islands, thereby thinning the insulating domain walls and increasing the probability of percolation.The evolution of the bubble morphology and the resulting suppression of global connectivity are detailed in Sec. C of the Supplemental Material . Therefore, to stabilize the intermediate glassy phase, we specifically fix $\alpha=1/16$ to maintain a sparse bubble density and restrict the pairing to a small value $\Delta/U=0.005$ to limit the bubble size, preventing the formation of a percolating cluster.

From a macroscopic perspective, the localized nature of the bubbles implies that the system, while globally insulating, retains local superfluid susceptibility. Figure~\ref{fig:eos}(a) shows the equation of state. While the Mott phase is characterized by an incompressible plateau ($\kappa = \partial n / \partial \mu = 0$), the Bubble Phase exhibits a continuously varying density with a finite slope, indicating a finite compressibility $\kappa > 0$. To make this explicit, we analyze the on-site particle number fluctuations $\langle \delta \hat{n}^2 \rangle = \langle \hat{n}^2 \rangle - \langle \hat{n} \rangle^2$ in Fig.~\ref{fig:eos}(b). Based on the fluctuation-dissipation relation, the compressibility is linked to local density fluctuations. As shown in Fig.\ref{fig:eos}(b), while fluctuations are suppressed to zero in the Mott limit, the Bubble Phase exhibits a significant finite value. This confirms that within the superfluid bubbles, particles are delocalized and compressible, even if transport is blocked globally.

The coexistence of finite compressibility (locally superfluid) and global insulating behavior (lack of percolation due to domain walls) is the defining phenomenology of a Bose Glass. Unlike conventional Bose glasses induced by extrinsic randomness, here the glassy state emerges in a disorder-free setting, driven purely by deterministic phase frustration. We thus identify this state as an Intrinsic Bose Glass.

\prlsection{Discussion}The emergence of the Bubble Phase in our frustrated Hamiltonian is not an isolated phenomenon but rather a specific manifestation of a generic mechanism driving frustration-induced localization. While the coexistence of superfluid clusters and insulating barriers is conventionally attributed to quenched disorder~\cite{fisher_boson_1989,sengupta_quantum_2007,piekarska_quantum_2018}, our work demonstrates that it can be generically understood through a unified framework of phase frustration. This framework bridges the gap between disorder-driven phenomena and similar fragmented states arising from deterministic inhomogeneities~\cite{hettiarachchilage_phase_2013,cruz_bose-hubbard_2014,zhang_dipolar_2025,zhang_quantum_2023}.

To demystify the physical origin of this frustration, consider a generic bosonic system where the hopping amplitude $t_{ij}$ is spatially modulated. Any such spatial texture $t(\bm{R}_{ij})$ can be decomposed via Fourier analysis into a spectrum of spatial frequencies:
$t_{ij} = \sum_{\bm{q}} \mathcal{T}_{\bm{q}} \, e^{i \bm{q} \cdot \bm{R}_{ij}}$.
This decomposition offers a direct mapping to the standard Peierls substitution, identifying an effective gauge potential $\bm{A}_{\text{eff}}(\bm{R}_{ij}) \equiv (\bm{q} \cdot \bm{R}_{ij}) \, \boldsymbol{\delta}$. Since $\bm{A}_{\text{eff}}(\bm{R})$ is linear in coordinates, it corresponds physically to a uniform magnetic field. Consequently, an arbitrary hopping texture involving multiple $\bm{q}$ modes is equivalent to subjecting the boson to a superposition of distinct, competing magnetic fields, leading to severe phase frustration. To resolve this tension, the system spontaneously creates insulating domain walls to screen the conflicting phase gradients, resulting in the localized bubbles.

This perspective offers a fresh insight into the nature of the Bose Glass caused by off-diagonal disorder. Traditionally, the Bose Glass is viewed as a consequence of Anderson localization induced by diagonal disorder~\cite{fisher_boson_1989}. However, in the context of off-diagonal disorder~\cite{piekarska_quantum_2018}, a stochastic distribution of hopping strengths corresponds to a broad, continuous spectrum of $\mathbf{q}$ modes. Physically, this represents the limit of ``extreme phase frustration.'' Unable to satisfy this chaotic set of gauge-imposed constraints, the system's global phase coherence shatters. Our ``Intrinsic Bose Glass,'' generated here by deterministic phase frustration, thus serves as a pristine, disorder-free archetype of this universal mechanism.

\section*{Data Availability Statement}
The code and data used to generate the figures in this study are available from the corresponding author upon reasonable request.


\bibliographystyle{apsrev4-2} 

\begin{thebibliography}{28}%
\makeatletter
\providecommand \@ifxundefined [1]{%
 \@ifx{#1\undefined}
}%
\providecommand \@ifnum [1]{%
 \ifnum #1\expandafter \@firstoftwo
 \else \expandafter \@secondoftwo
 \fi
}%
\providecommand \@ifx [1]{%
 \ifx #1\expandafter \@firstoftwo
 \else \expandafter \@secondoftwo
 \fi
}%
\providecommand \natexlab [1]{#1}%
\providecommand \enquote  [1]{``#1''}%
\providecommand \bibnamefont  [1]{#1}%
\providecommand \bibfnamefont [1]{#1}%
\providecommand \citenamefont [1]{#1}%
\providecommand \href@noop [0]{\@secondoftwo}%
\providecommand \href [0]{\begingroup \@sanitize@url \@href}%
\providecommand \@href[1]{\@@startlink{#1}\@@href}%
\providecommand \@@href[1]{\endgroup#1\@@endlink}%
\providecommand \@sanitize@url [0]{\catcode `\\12\catcode `\$12\catcode `\&12\catcode `\#12\catcode `\^12\catcode `\_12\catcode `\%12\relax}%
\providecommand \@@startlink[1]{}%
\providecommand \@@endlink[0]{}%
\providecommand \url  [0]{\begingroup\@sanitize@url \@url }%
\providecommand \@url [1]{\endgroup\@href {#1}{\urlprefix }}%
\providecommand \urlprefix  [0]{URL }%
\providecommand \Eprint [0]{\href }%
\providecommand \doibase [0]{https://doi.org/}%
\providecommand \selectlanguage [0]{\@gobble}%
\providecommand \bibinfo  [0]{\@secondoftwo}%
\providecommand \bibfield  [0]{\@secondoftwo}%
\providecommand \translation [1]{[#1]}%
\providecommand \BibitemOpen [0]{}%
\providecommand \bibitemStop [0]{}%
\providecommand \bibitemNoStop [0]{.\EOS\space}%
\providecommand \EOS [0]{\spacefactor3000\relax}%
\providecommand \BibitemShut  [1]{\csname bibitem#1\endcsname}%
\let\auto@bib@innerbib\@empty
\bibitem [{\citenamefont {Anderson}(1958)}]{anderson_absence_1958}%
  \BibitemOpen
  \bibfield  {author} {\bibinfo {author} {\bibfnamefont {P.~W.}\ \bibnamefont {Anderson}},\ }\href {https://doi.org/10.1103/PhysRev.109.1492} {\bibfield  {journal} {\bibinfo  {journal} {Phys. Rev.}\ }\textbf {\bibinfo {volume} {109}},\ \bibinfo {pages} {1492} (\bibinfo {year} {1958})}\BibitemShut {NoStop}%
\bibitem [{\citenamefont {Fisher}\ \emph {et~al.}(1989)\citenamefont {Fisher}, \citenamefont {Weichman}, \citenamefont {Grinstein},\ and\ \citenamefont {Fisher}}]{fisher_boson_1989}%
  \BibitemOpen
  \bibfield  {author} {\bibinfo {author} {\bibfnamefont {M.~P.~A.}\ \bibnamefont {Fisher}}, \bibinfo {author} {\bibfnamefont {P.~B.}\ \bibnamefont {Weichman}}, \bibinfo {author} {\bibfnamefont {G.}~\bibnamefont {Grinstein}},\ and\ \bibinfo {author} {\bibfnamefont {D.~S.}\ \bibnamefont {Fisher}},\ }\href {https://doi.org/10.1103/PhysRevB.40.546} {\bibfield  {journal} {\bibinfo  {journal} {Phys. Rev. B}\ }\textbf {\bibinfo {volume} {40}},\ \bibinfo {pages} {546} (\bibinfo {year} {1989})}\BibitemShut {NoStop}%
\bibitem [{\citenamefont {Fallani}\ \emph {et~al.}(2007)\citenamefont {Fallani}, \citenamefont {Lye}, \citenamefont {Guarrera}, \citenamefont {Fort},\ and\ \citenamefont {Inguscio}}]{fallani_ultracold_2007}%
  \BibitemOpen
  \bibfield  {author} {\bibinfo {author} {\bibfnamefont {L.}~\bibnamefont {Fallani}}, \bibinfo {author} {\bibfnamefont {J.~E.}\ \bibnamefont {Lye}}, \bibinfo {author} {\bibfnamefont {V.}~\bibnamefont {Guarrera}}, \bibinfo {author} {\bibfnamefont {C.}~\bibnamefont {Fort}},\ and\ \bibinfo {author} {\bibfnamefont {M.}~\bibnamefont {Inguscio}},\ }\href {https://doi.org/10.1103/PhysRevLett.98.130404} {\bibfield  {journal} {\bibinfo  {journal} {Phys. Rev. Lett.}\ }\textbf {\bibinfo {volume} {98}},\ \bibinfo {pages} {130404} (\bibinfo {year} {2007})}\BibitemShut {NoStop}%
\bibitem [{\citenamefont {Sengupta}\ and\ \citenamefont {Haas}(2007)}]{sengupta_quantum_2007}%
  \BibitemOpen
  \bibfield  {author} {\bibinfo {author} {\bibfnamefont {P.}~\bibnamefont {Sengupta}}\ and\ \bibinfo {author} {\bibfnamefont {S.}~\bibnamefont {Haas}},\ }\href {https://doi.org/10.1103/PhysRevLett.99.050403} {\bibfield  {journal} {\bibinfo  {journal} {Phys. Rev. Lett.}\ }\textbf {\bibinfo {volume} {99}},\ \bibinfo {pages} {050403} (\bibinfo {year} {2007})}\BibitemShut {NoStop}%
\bibitem [{\citenamefont {Piekarska}\ and\ \citenamefont {Kope{\'c}}(2018)}]{piekarska_quantum_2018}%
  \BibitemOpen
  \bibfield  {author} {\bibinfo {author} {\bibfnamefont {A.~M.}\ \bibnamefont {Piekarska}}\ and\ \bibinfo {author} {\bibfnamefont {T.~K.}\ \bibnamefont {Kope{\'c}}},\ }\href {https://doi.org/10.1103/PhysRevLett.120.160401} {\bibfield  {journal} {\bibinfo  {journal} {Phys. Rev. Lett.}\ }\textbf {\bibinfo {volume} {120}},\ \bibinfo {pages} {160401} (\bibinfo {year} {2018})}\BibitemShut {NoStop}%
\bibitem [{\citenamefont {Piekarska}\ and\ \citenamefont {Kope{\'c}}(2022)}]{piekarska_emergence_2022}%
  \BibitemOpen
  \bibfield  {author} {\bibinfo {author} {\bibfnamefont {A.~M.}\ \bibnamefont {Piekarska}}\ and\ \bibinfo {author} {\bibfnamefont {T.~K.}\ \bibnamefont {Kope{\'c}}},\ }\href {https://doi.org/10.1103/PhysRevB.105.174203} {\bibfield  {journal} {\bibinfo  {journal} {Phys. Rev. B}\ }\textbf {\bibinfo {volume} {105}},\ \bibinfo {pages} {174203} (\bibinfo {year} {2022})}\BibitemShut {NoStop}%
\bibitem [{\citenamefont {Smith}\ \emph {et~al.}(2017)\citenamefont {Smith}, \citenamefont {Knolle}, \citenamefont {Kovrizhin},\ and\ \citenamefont {Moessner}}]{smith_disorder-free_2017}%
  \BibitemOpen
  \bibfield  {author} {\bibinfo {author} {\bibfnamefont {A.}~\bibnamefont {Smith}}, \bibinfo {author} {\bibfnamefont {J.}~\bibnamefont {Knolle}}, \bibinfo {author} {\bibfnamefont {D.~L.}\ \bibnamefont {Kovrizhin}},\ and\ \bibinfo {author} {\bibfnamefont {R.}~\bibnamefont {Moessner}},\ }\href {https://doi.org/10.1103/PhysRevLett.118.266601} {\bibfield  {journal} {\bibinfo  {journal} {Phys. Rev. Lett.}\ }\textbf {\bibinfo {volume} {118}},\ \bibinfo {pages} {266601} (\bibinfo {year} {2017})}\BibitemShut {NoStop}%
\bibitem [{\citenamefont {Karpov}\ \emph {et~al.}(2021)\citenamefont {Karpov}, \citenamefont {Verdel}, \citenamefont {Huang}, \citenamefont {Schmitt},\ and\ \citenamefont {Heyl}}]{karpov_disorder-free_2021}%
  \BibitemOpen
  \bibfield  {author} {\bibinfo {author} {\bibfnamefont {P.}~\bibnamefont {Karpov}}, \bibinfo {author} {\bibfnamefont {R.}~\bibnamefont {Verdel}}, \bibinfo {author} {\bibfnamefont {Y.-P.}\ \bibnamefont {Huang}}, \bibinfo {author} {\bibfnamefont {M.}~\bibnamefont {Schmitt}},\ and\ \bibinfo {author} {\bibfnamefont {M.}~\bibnamefont {Heyl}},\ }\href {https://doi.org/10.1103/PhysRevLett.126.130401} {\bibfield  {journal} {\bibinfo  {journal} {Phys. Rev. Lett.}\ }\textbf {\bibinfo {volume} {126}},\ \bibinfo {pages} {130401} (\bibinfo {year} {2021})}\BibitemShut {NoStop}%
\bibitem [{\citenamefont {Aubry}\ and\ \citenamefont {Andr{\'e}}(1980)}]{aubry_analyticity_1980}%
  \BibitemOpen
  \bibfield  {author} {\bibinfo {author} {\bibfnamefont {S.}~\bibnamefont {Aubry}}\ and\ \bibinfo {author} {\bibfnamefont {G.}~\bibnamefont {Andr{\'e}}},\ }\href@noop {} {\bibfield  {journal} {\bibinfo  {journal} {Ann. Israel Phys. Soc.}\ }\textbf {\bibinfo {volume} {3}},\ \bibinfo {pages} {133} (\bibinfo {year} {1980})}\BibitemShut {NoStop}%
\bibitem [{\citenamefont {Roscilde}(2008)}]{roscilde_bosons_2008}%
  \BibitemOpen
  \bibfield  {author} {\bibinfo {author} {\bibfnamefont {T.}~\bibnamefont {Roscilde}},\ }\href {https://doi.org/10.1103/PhysRevA.77.063605} {\bibfield  {journal} {\bibinfo  {journal} {Phys. Rev. A}\ }\textbf {\bibinfo {volume} {77}},\ \bibinfo {pages} {063605} (\bibinfo {year} {2008})}\BibitemShut {NoStop}%
\bibitem [{\citenamefont {Ding}\ \emph {et~al.}(2025)\citenamefont {Ding}, \citenamefont {Lang}, \citenamefont {Zhu},\ and\ \citenamefont {He}}]{ding_interaction-induced_2025}%
  \BibitemOpen
  \bibfield  {author} {\bibinfo {author} {\bibfnamefont {S.-H.}\ \bibnamefont {Ding}}, \bibinfo {author} {\bibfnamefont {L.-J.}\ \bibnamefont {Lang}}, \bibinfo {author} {\bibfnamefont {Q.}~\bibnamefont {Zhu}},\ and\ \bibinfo {author} {\bibfnamefont {L.}~\bibnamefont {He}},\ }\href {https://doi.org/10.1103/fvny-58kf} {\bibfield  {journal} {\bibinfo  {journal} {Phys. Rev. A}\ }\textbf {\bibinfo {volume} {112}},\ \bibinfo {pages} {033322} (\bibinfo {year} {2025})}\BibitemShut {NoStop}%
\bibitem [{\citenamefont {Zhang}\ \emph {et~al.}(2025)\citenamefont {Zhang}, \citenamefont {Fan}, \citenamefont {{Capogrosso-Sansone}},\ and\ \citenamefont {Deng}}]{zhang_dipolar_2025}%
  \BibitemOpen
  \bibfield  {author} {\bibinfo {author} {\bibfnamefont {C.}~\bibnamefont {Zhang}}, \bibinfo {author} {\bibfnamefont {Z.}~\bibnamefont {Fan}}, \bibinfo {author} {\bibfnamefont {B.}~\bibnamefont {{Capogrosso-Sansone}}},\ and\ \bibinfo {author} {\bibfnamefont {Y.}~\bibnamefont {Deng}},\ }\href {https://doi.org/10.1103/PhysRevB.111.024511} {\bibfield  {journal} {\bibinfo  {journal} {Phys. Rev. B}\ }\textbf {\bibinfo {volume} {111}},\ \bibinfo {pages} {024511} (\bibinfo {year} {2025})}\BibitemShut {NoStop}%
\bibitem [{\citenamefont {Ciardi}\ \emph {et~al.}(2023)\citenamefont {Ciardi}, \citenamefont {Angelone}, \citenamefont {Mezzacapo},\ and\ \citenamefont {Cinti}}]{ciardi_quasicrystalline_2023}%
  \BibitemOpen
  \bibfield  {author} {\bibinfo {author} {\bibfnamefont {M.}~\bibnamefont {Ciardi}}, \bibinfo {author} {\bibfnamefont {A.}~\bibnamefont {Angelone}}, \bibinfo {author} {\bibfnamefont {F.}~\bibnamefont {Mezzacapo}},\ and\ \bibinfo {author} {\bibfnamefont {F.}~\bibnamefont {Cinti}},\ }\href {https://doi.org/10.1103/PhysRevLett.131.173402} {\bibfield  {journal} {\bibinfo  {journal} {Phys. Rev. Lett.}\ }\textbf {\bibinfo {volume} {131}},\ \bibinfo {pages} {173402} (\bibinfo {year} {2023})}\BibitemShut {NoStop}%
\bibitem [{\citenamefont {Powell}\ \emph {et~al.}(2011)\citenamefont {Powell}, \citenamefont {Barnett}, \citenamefont {Sensarma},\ and\ \citenamefont {Sarma}}]{powell_bogoliubov_2011}%
  \BibitemOpen
  \bibfield  {author} {\bibinfo {author} {\bibfnamefont {S.}~\bibnamefont {Powell}}, \bibinfo {author} {\bibfnamefont {R.}~\bibnamefont {Barnett}}, \bibinfo {author} {\bibfnamefont {R.}~\bibnamefont {Sensarma}},\ and\ \bibinfo {author} {\bibfnamefont {S.~D.}\ \bibnamefont {Sarma}},\ }\href {https://doi.org/10.1103/PhysRevA.83.013612} {\bibfield  {journal} {\bibinfo  {journal} {Phys. Rev. A}\ }\textbf {\bibinfo {volume} {83}},\ \bibinfo {pages} {013612} (\bibinfo {year} {2011})}\BibitemShut {NoStop}%
\bibitem [{\citenamefont {Sachdeva}\ and\ \citenamefont {Ghosh}(2012)}]{sachdeva_density_2012}%
  \BibitemOpen
  \bibfield  {author} {\bibinfo {author} {\bibfnamefont {R.}~\bibnamefont {Sachdeva}}\ and\ \bibinfo {author} {\bibfnamefont {S.}~\bibnamefont {Ghosh}},\ }\href {https://doi.org/10.1103/PhysRevA.85.013642} {\bibfield  {journal} {\bibinfo  {journal} {Phys. Rev. A}\ }\textbf {\bibinfo {volume} {85}},\ \bibinfo {pages} {013642} (\bibinfo {year} {2012})}\BibitemShut {NoStop}%
\bibitem [{\citenamefont {Dhar}\ \emph {et~al.}(2012)\citenamefont {Dhar}, \citenamefont {Maji}, \citenamefont {Mishra}, \citenamefont {Pai}, \citenamefont {Mukerjee},\ and\ \citenamefont {Paramekanti}}]{dhar_bose-hubbard_2012}%
  \BibitemOpen
  \bibfield  {author} {\bibinfo {author} {\bibfnamefont {A.}~\bibnamefont {Dhar}}, \bibinfo {author} {\bibfnamefont {M.}~\bibnamefont {Maji}}, \bibinfo {author} {\bibfnamefont {T.}~\bibnamefont {Mishra}}, \bibinfo {author} {\bibfnamefont {R.~V.}\ \bibnamefont {Pai}}, \bibinfo {author} {\bibfnamefont {S.}~\bibnamefont {Mukerjee}},\ and\ \bibinfo {author} {\bibfnamefont {A.}~\bibnamefont {Paramekanti}},\ }\href {https://doi.org/10.1103/PhysRevA.85.041602} {\bibfield  {journal} {\bibinfo  {journal} {Phys. Rev. A}\ }\textbf {\bibinfo {volume} {85}},\ \bibinfo {pages} {041602} (\bibinfo {year} {2012})}\BibitemShut {NoStop}%
\bibitem [{\citenamefont {Zhou}\ \emph {et~al.}(2011)\citenamefont {Zhou}, \citenamefont {Porto},\ and\ \citenamefont {{Das Sarma}}}]{zhou_condensates_2011}%
  \BibitemOpen
  \bibfield  {author} {\bibinfo {author} {\bibfnamefont {Q.}~\bibnamefont {Zhou}}, \bibinfo {author} {\bibfnamefont {J.~V.}\ \bibnamefont {Porto}},\ and\ \bibinfo {author} {\bibfnamefont {S.}~\bibnamefont {{Das Sarma}}},\ }\href {https://doi.org/10.1103/PhysRevB.83.195106} {\bibfield  {journal} {\bibinfo  {journal} {Phys. Rev. B}\ }\textbf {\bibinfo {volume} {83}},\ \bibinfo {pages} {195106} (\bibinfo {year} {2011})}\BibitemShut {NoStop}
  \bibitem [{\citenamefont {Slim}\ \emph {et~al.}(2024)\citenamefont {Slim}, \citenamefont {Wanjura}, \citenamefont {Brunelli}, \citenamefont {del Pino}, \citenamefont {Nunnenkamp},\ and\ \citenamefont {Verhagen}}]{slim_optomechanical_2024}%
  \BibitemOpen
  \bibfield  {author} {\bibinfo {author} {\bibfnamefont {J.~J.}\ \bibnamefont {Slim}}, \bibinfo {author} {\bibfnamefont {C.~C.}\ \bibnamefont {Wanjura}}, \bibinfo {author} {\bibfnamefont {M.}~\bibnamefont {Brunelli}}, \bibinfo {author} {\bibfnamefont {J.}~\bibnamefont {del Pino}}, \bibinfo {author} {\bibfnamefont {A.}~\bibnamefont {Nunnenkamp}},\ and\ \bibinfo {author} {\bibfnamefont {E.}~\bibnamefont {Verhagen}},\ }\href {https://doi.org/10.1038/s41586-024-07174-w} {\bibfield  {journal} {\bibinfo  {journal} {Nature}\ }\textbf {\bibinfo {volume} {627}},\ \bibinfo {pages} {767} (\bibinfo {year} {2024})}\BibitemShut {NoStop}%
\bibitem [{\citenamefont {Busnaina}\ \emph {et~al.}(2024)\citenamefont {Busnaina}, \citenamefont {Shi}, \citenamefont {McDonald}, \citenamefont {Dubyna}, \citenamefont {Nsanzineza}, \citenamefont {Hung}, \citenamefont {Chang}, \citenamefont {Clerk},\ and\ \citenamefont {Wilson}}]{busnaina_quantum_2024}%
  \BibitemOpen
  \bibfield  {author} {\bibinfo {author} {\bibfnamefont {J.~H.}\ \bibnamefont {Busnaina}}, \bibinfo {author} {\bibfnamefont {Z.}~\bibnamefont {Shi}}, \bibinfo {author} {\bibfnamefont {A.}~\bibnamefont {McDonald}}, \bibinfo {author} {\bibfnamefont {D.}~\bibnamefont {Dubyna}}, \bibinfo {author} {\bibfnamefont {I.}~\bibnamefont {Nsanzineza}}, \bibinfo {author} {\bibfnamefont {J.~S.~C.}\ \bibnamefont {Hung}}, \bibinfo {author} {\bibfnamefont {C.~W.~S.}\ \bibnamefont {Chang}}, \bibinfo {author} {\bibfnamefont {A.~A.}\ \bibnamefont {Clerk}},\ and\ \bibinfo {author} {\bibfnamefont {C.~M.}\ \bibnamefont {Wilson}},\ }\href {https://doi.org/10.1038/s41467-024-47186-8} {\bibfield  {journal} {\bibinfo  {journal} {Nat. Commun.}\ }\textbf {\bibinfo {volume} {15}},\ \bibinfo {pages} {3065} (\bibinfo {year} {2024})}\BibitemShut {NoStop}%
\bibitem [{\citenamefont {McDonald}\ \emph {et~al.}(2018)\citenamefont {McDonald}, \citenamefont {{Pereg-Barnea}},\ and\ \citenamefont {Clerk}}]{mcdonald_phase-dependent_2018}%
  \BibitemOpen
  \bibfield  {author} {\bibinfo {author} {\bibfnamefont {A.}~\bibnamefont {McDonald}}, \bibinfo {author} {\bibfnamefont {T.}~\bibnamefont {{Pereg-Barnea}}},\ and\ \bibinfo {author} {\bibfnamefont {A.~A.}\ \bibnamefont {Clerk}},\ }\href {https://doi.org/10.1103/PhysRevX.8.041031} {\bibfield  {journal} {\bibinfo  {journal} {Phys. Rev. X}\ }\textbf {\bibinfo {volume} {8}},\ \bibinfo {pages} {041031} (\bibinfo {year} {2018})}\BibitemShut {NoStop}%
\bibitem [{\citenamefont {Vishveshwara}\ and\ \citenamefont {Weld}(2021)}]{vishveshwara_z2_2021}%
  \BibitemOpen
  \bibfield  {author} {\bibinfo {author} {\bibfnamefont {S.}~\bibnamefont {Vishveshwara}}\ and\ \bibinfo {author} {\bibfnamefont {D.~M.}\ \bibnamefont {Weld}},\ }\href {https://doi.org/10.1103/PhysRevA.103.L051301} {\bibfield  {journal} {\bibinfo  {journal} {Phys. Rev. A}\ }\textbf {\bibinfo {volume} {103}},\ \bibinfo {pages} {L051301} (\bibinfo {year} {2021})}\BibitemShut {NoStop}%
\bibitem [{\citenamefont {Wang}\ \emph {et~al.}(2022)\citenamefont {Wang}, \citenamefont {You},\ and\ \citenamefont {Sun}}]{wang_quantum_2022}%
  \BibitemOpen
  \bibfield  {author} {\bibinfo {author} {\bibfnamefont {Y.-N.}\ \bibnamefont {Wang}}, \bibinfo {author} {\bibfnamefont {W.-L.}\ \bibnamefont {You}},\ and\ \bibinfo {author} {\bibfnamefont {G.}~\bibnamefont {Sun}},\ }\href {https://doi.org/10.1103/PhysRevA.106.053315} {\bibfield  {journal} {\bibinfo  {journal} {Phys. Rev. A}\ }\textbf {\bibinfo {volume} {106}},\ \bibinfo {pages} {053315} (\bibinfo {year} {2022})}\BibitemShut {NoStop}%
\bibitem [{\citenamefont {Peotta}\ \emph {et~al.}(2014)\citenamefont {Peotta}, \citenamefont {Mazza}, \citenamefont {Vicari}, \citenamefont {Polini}, \citenamefont {Fazio},\ and\ \citenamefont {Rossini}}]{peotta_xyz_2014}%
  \BibitemOpen
  \bibfield  {author} {\bibinfo {author} {\bibfnamefont {S.}~\bibnamefont {Peotta}}, \bibinfo {author} {\bibfnamefont {L.}~\bibnamefont {Mazza}}, \bibinfo {author} {\bibfnamefont {E.}~\bibnamefont {Vicari}}, \bibinfo {author} {\bibfnamefont {M.}~\bibnamefont {Polini}}, \bibinfo {author} {\bibfnamefont {R.}~\bibnamefont {Fazio}},\ and\ \bibinfo {author} {\bibfnamefont {D.}~\bibnamefont {Rossini}},\ }\href {https://doi.org/10.1088/1742-5468/2014/09/P09005} {\bibfield  {journal} {\bibinfo  {journal} {J. Stat. Mech.}\ }\textbf {\bibinfo {volume} {2014}},\ \bibinfo {pages} {P09005} (\bibinfo {year} {2014})}\BibitemShut {NoStop}%
\bibitem [{\citenamefont {Buonsante}\ \emph {et~al.}(2009)\citenamefont {Buonsante}, \citenamefont {Massel}, \citenamefont {Penna},\ and\ \citenamefont {Vezzani}}]{buonsante_gutzwiller_2009}%
  \BibitemOpen
  \bibfield  {author} {\bibinfo {author} {\bibfnamefont {P.}~\bibnamefont {Buonsante}}, \bibinfo {author} {\bibfnamefont {F.}~\bibnamefont {Massel}}, \bibinfo {author} {\bibfnamefont {V.}~\bibnamefont {Penna}},\ and\ \bibinfo {author} {\bibfnamefont {A.}~\bibnamefont {Vezzani}},\ }\href {https://doi.org/10.1103/PhysRevA.79.013623} {\bibfield  {journal} {\bibinfo  {journal} {Phys. Rev. A}\ }\textbf {\bibinfo {volume} {79}},\ \bibinfo {pages} {013623} (\bibinfo {year} {2009})}\BibitemShut {NoStop}%
\bibitem [{\citenamefont {Suthar}\ \emph {et~al.}(2020)\citenamefont {Suthar}, \citenamefont {Sable}, \citenamefont {Bai}, \citenamefont {Bandyopadhyay}, \citenamefont {Pal},\ and\ \citenamefont {Angom}}]{suthar_supersolid_2020}%
  \BibitemOpen
  \bibfield  {author} {\bibinfo {author} {\bibfnamefont {K.}~\bibnamefont {Suthar}}, \bibinfo {author} {\bibfnamefont {H.}~\bibnamefont {Sable}}, \bibinfo {author} {\bibfnamefont {R.}~\bibnamefont {Bai}}, \bibinfo {author} {\bibfnamefont {S.}~\bibnamefont {Bandyopadhyay}}, \bibinfo {author} {\bibfnamefont {S.}~\bibnamefont {Pal}},\ and\ \bibinfo {author} {\bibfnamefont {D.}~\bibnamefont {Angom}},\ }\href {https://doi.org/10.1103/PhysRevA.102.013320} {\bibfield  {journal} {\bibinfo  {journal} {Phys. Rev. A}\ }\textbf {\bibinfo {volume} {102}},\ \bibinfo {pages} {013320} (\bibinfo {year} {2020})}\BibitemShut {NoStop}%
\bibitem [{\citenamefont {Hoshen}\ and\ \citenamefont {Kopelman}(1976)}]{hoshen_percolation_1976}%
  \BibitemOpen
  \bibfield  {author} {\bibinfo {author} {\bibfnamefont {J.}~\bibnamefont {Hoshen}}\ and\ \bibinfo {author} {\bibfnamefont {R.}~\bibnamefont {Kopelman}},\ }\href {https://doi.org/10.1103/PhysRevB.14.3438} {\bibfield  {journal} {\bibinfo  {journal} {Phys. Rev. B}\ }\textbf {\bibinfo {volume} {14}},\ \bibinfo {pages} {3438} (\bibinfo {year} {1976})}\BibitemShut {NoStop}%
\bibitem [{\citenamefont {Caleffi}\ \emph {et~al.}(2020)\citenamefont {Caleffi}, \citenamefont {Capone}, \citenamefont {Menotti}, \citenamefont {Carusotto},\ and\ \citenamefont {Recati}}]{caleffi_quantum_2020}%
  \BibitemOpen
  \bibfield  {author} {\bibinfo {author} {\bibfnamefont {F.}~\bibnamefont {Caleffi}}, \bibinfo {author} {\bibfnamefont {M.}~\bibnamefont {Capone}}, \bibinfo {author} {\bibfnamefont {C.}~\bibnamefont {Menotti}}, \bibinfo {author} {\bibfnamefont {I.}~\bibnamefont {Carusotto}},\ and\ \bibinfo {author} {\bibfnamefont {A.}~\bibnamefont {Recati}},\ }\href {https://doi.org/10.1103/PhysRevResearch.2.033276} {\bibfield  {journal} {\bibinfo  {journal} {Phys. Rev. Research}\ }\textbf {\bibinfo {volume} {2}},\ \bibinfo {pages} {033276} (\bibinfo {year} {2020})}\BibitemShut {NoStop}%
\bibitem [{\citenamefont {Hettiarachchilage}\ \emph {et~al.}(2013)\citenamefont {Hettiarachchilage}, \citenamefont {Rousseau}, \citenamefont {Tam}, \citenamefont {Jarrell},\ and\ \citenamefont {Moreno}}]{hettiarachchilage_phase_2013}%
  \BibitemOpen
  \bibfield  {author} {\bibinfo {author} {\bibfnamefont {K.}~\bibnamefont {Hettiarachchilage}}, \bibinfo {author} {\bibfnamefont {V.~G.}\ \bibnamefont {Rousseau}}, \bibinfo {author} {\bibfnamefont {K.-M.}\ \bibnamefont {Tam}}, \bibinfo {author} {\bibfnamefont {M.}~\bibnamefont {Jarrell}},\ and\ \bibinfo {author} {\bibfnamefont {J.}~\bibnamefont {Moreno}},\ }\href {https://doi.org/10.1103/PhysRevA.87.051607} {\bibfield  {journal} {\bibinfo  {journal} {Phys. Rev. A}\ }\textbf {\bibinfo {volume} {87}},\ \bibinfo {pages} {051607} (\bibinfo {year} {2013})}\BibitemShut {NoStop}%
\bibitem [{\citenamefont {Cruz}\ \emph {et~al.}(2014)\citenamefont {Cruz}, \citenamefont {Franco},\ and\ \citenamefont {{Silva-Valencia}}}]{cruz_bose-hubbard_2014}%
  \BibitemOpen
  \bibfield  {author} {\bibinfo {author} {\bibfnamefont {G.~J.}\ \bibnamefont {Cruz}}, \bibinfo {author} {\bibfnamefont {R.}~\bibnamefont {Franco}},\ and\ \bibinfo {author} {\bibfnamefont {J.}~\bibnamefont {{Silva-Valencia}}},\ }\href {https://doi.org/10.1088/1742-6596/480/1/012003} {\bibfield  {journal} {\bibinfo  {journal} {J. Phys.: Conf. Ser.}\ }\textbf {\bibinfo {volume} {480}},\ \bibinfo {pages} {012003} (\bibinfo {year} {2014})}\BibitemShut {NoStop}%
\bibitem [{\citenamefont {Zhang}\ and\ \citenamefont {Yang}(2023)}]{zhang_quantum_2023}%
  \BibitemOpen
  \bibfield  {author} {\bibinfo {author} {\bibfnamefont {X.-r.}\ \bibnamefont {Zhang}}\ and\ \bibinfo {author} {\bibfnamefont {S.-J.}\ \bibnamefont {Yang}},\ }\href {https://doi.org/10.1016/j.rinp.2023.106998} {\bibfield  {journal} {\bibinfo  {journal} {Results Phys.}\ }\textbf {\bibinfo {volume} {53}},\ \bibinfo {pages} {106998} (\bibinfo {year} {2023})}\BibitemShut {NoStop}%
\end{thebibliography}
%

\clearpage
\onecolumngrid
\appendix

\setcounter{equation}{0}
\setcounter{figure}{0}
\setcounter{table}{0}
\setcounter{page}{1}
\renewcommand{\theequation}{S\arabic{equation}}
\renewcommand{\thefigure}{S\arabic{figure}}
\renewcommand{\thetable}{S\arabic{table}}
\renewcommand{\thesection}{\Alph{section}}

\begin{center}
\textbf{\large Supplementary Material for: \\ Phase Frustration Induced Intrinsic Bose Glass in the Kitaev-Bose-Hubbard Model}
\end{center}

\section{A. Explicit Form of the BdG Dynamical Matrix in Slave-Particle Formalism}

In this section, we provide the explicit expressions for the matrix elements of the Bogoliubov-de Gennes (BdG) dynamical matrix $\mathcal{M}_{\text{BdG}}$ derived within the slave-particle formalism. As introduced in the main text, the fluctuation operators are defined via the slave particles $\hat{c}_{n,i}$, where $\hat{b}_i = \sum_n \sqrt{n} \hat{c}_{n,i}^\dagger \hat{c}_{n-1,i}$. The BdG Hamiltonian in the Nambu basis $\delta\hat{\Phi} = (\delta\hat{\mathbf{c}}, \delta\hat{\mathbf{c}}^\dagger)^T$ takes the block form:
\begin{equation}
\mathcal{M}_{\text{BdG}} = 
\begin{pmatrix}
\mathbf{A} & \mathbf{B} \\
-\mathbf{B}^* & -\mathbf{A}^*
\end{pmatrix}.
\end{equation}
Here, the indices of the submatrices run over $(i,m)$ and $(j,n)$, representing the $m$-th Fock state at site $i$ and the $n$-th Fock state at site $j$, respectively. The hopping amplitude is denoted by $J$, the pairing amplitude by $\Delta$, and the Peierls phase by $\theta_{ij}$. We assume the pairing phase follows the gauge field ($\phi_{ij} = \theta_{ij}$). The mean-field parameters are defined based on the local ground state coefficients $c_{n,i}^{(0)}$:
\begin{align}
\psi_i &= \sum_n \sqrt{n} c_{n-1,i}^{(0)*} c_{n,i}^{(0)}, \nonumber \\
\bar{n}_i &= \sum_n n |c_{n,i}^{(0)}|^2, \nonumber \\
\overline{n(n-1)}_i &= \sum_n n(n-1) |c_{n,i}^{(0)}|^2.
\end{align}

\subsection{1. Matrix A (Normal Terms)}
The matrix $\mathbf{A}$ contains diagonal and off-diagonal terms associated with particle number conservation (in the hopping sector) and anomalous mixing (in the pairing sector).

\textbf{On-site Diagonal Elements} ($i=j, m=n$):
These terms describe the local energy cost including interaction, chemical potential, and the mean-field energy shift from neighbors:
\begin{equation}
\begin{aligned}
A_{im,im} &= \frac{U}{2}[m(m-1) - \overline{n(n-1)}_i] - \mu(m - \bar{n}_i) \\
&\quad + \sum_{j \in \text{NN}(i)} 2 \operatorname{Re} \left[ e^{i\theta_{ij}} (J \psi_j \psi_i^* + \Delta \psi_i^* \psi_j^*) \right].
\end{aligned}
\end{equation}

\textbf{On-site Off-diagonal Elements} ($i=j, n=m-1$):
These terms arise from the coupling to the mean-field order parameter:
\begin{equation}
\begin{aligned}
A_{i,m,m-1} &= -\sqrt{m} \sum_{j \in \text{NN}(i)} \Big[ J (e^{i\theta_{ij}}\psi_j + e^{-i\theta_{ij}}\psi_j^*) \\
&\quad + \Delta (e^{i\theta_{ij}}\psi_j^* + e^{-i\theta_{ij}}\psi_j) \Big].
\end{aligned}
\end{equation}
Note that Hermiticity of the block $\mathbf{A}$ implies $A_{i,m-1,m} = A_{i,m,m-1}^*$.

\textbf{Off-site Elements} ($i \neq j$):
These terms describe the direct transfer or pairing of fluctuations between sites:
\begin{equation}
\begin{aligned}
A_{im,jn} &= -J \Big[ e^{i\theta_{ij}}\sqrt{mn} c_{m-1,i}^{(0)} c_{n-1,j}^{(0)*} \\
&\quad + e^{-i\theta_{ij}}\sqrt{(m+1)(n+1)} c_{m+1,i}^{(0)} c_{n+1,j}^{(0)*} \Big] \\
&\quad - \Delta \Big[ e^{i\theta_{ij}}\sqrt{m(n+1)} c_{m-1,i}^{(0)} c_{n+1,j}^{(0)*} \\
&\quad + e^{-i\theta_{ij}}\sqrt{n(m+1)} c_{n-1,j}^{(0)*} c_{m+1,i}^{(0)} \Big].
\end{aligned}
\end{equation}

\subsection{2. Matrix B (Anomalous Terms)}
The matrix $\mathbf{B}$ describes the creation/annihilation of pairs of fluctuations.

\textbf{On-site Elements}:
\begin{equation}
B_{im,in} = 0.
\end{equation}

\textbf{Off-site Elements} ($i \neq j$):
\begin{equation}
\begin{aligned}
B_{im,jn} &= -J \Big[ e^{i\theta_{ij}}\sqrt{m(n+1)} c_{m-1,i}^{(0)} c_{n+1,j}^{(0)} \\
&\quad + e^{-i\theta_{ij}}\sqrt{n(m+1)} c_{n-1,j}^{(0)} c_{m+1,i}^{(0)} \Big] \\
&\quad - \Delta \Big[ e^{i\theta_{ij}}\sqrt{mn} c_{m-1,i}^{(0)} c_{n-1,j}^{(0)} \\
&\quad + e^{-i\theta_{ij}}\sqrt{(m+1)(n+1)} c_{m+1,i}^{(0)} c_{n+1,j}^{(0)} \Big].
\end{aligned}
\end{equation}

The eigenvalues $\omega_\nu$ of $\mathcal{M}_{\text{BdG}}$ yield the collective excitation spectrum. The presence of any complex eigenvalue with $\operatorname{Im}(\omega_\nu) \neq 0$ signals dynamical instability.

\section{B. Perturbative Justification of the Energy Penalty Method}

In the slave-particle formalism, the Hilbert space is enlarged, introducing a set of spurious zero-energy modes arising from the local $U(1)$ gauge symmetry and normalization constraints. These modes satisfy $\mathcal{M}_{\text{BdG}} |\Psi_{\text{spur}}\rangle = 0$.
To decouple them from the physical spectrum, we construct an effective Hamiltonian $\mathcal{H}_{\text{eff}} = \mathcal{M}_{\text{BdG}} + \Lambda \mathcal{P}_{\text{spur}}$, where $\mathcal{P}_{\text{spur}}$ is the projection operator onto the spurious subspace and $\Lambda$ is a large penalty parameter. Here, we provide a rigorous justification based on standard perturbation theory (specifically, the projection technique for effective Hamiltonians). We demonstrate that, within a perturbative expansion in powers of $1/\Lambda$, this method effectively isolates physical modes with a vanishingly small energy shift, even in the presence of non-Hermitian couplings.

\subsection{1. Block Hamiltonian Formalism}

Let $\hat{P}$ and $\hat{Q}$ be the projection operators onto the physical subspace and the spurious subspace, respectively, satisfying $\hat{P} + \hat{Q} = \hat{1}$, $\hat{P}\hat{Q}=0$, and importantly $\hat{Q} \equiv \mathcal{P}_{\text{spur}}$. The eigenvalue problem $\mathcal{H}_{\text{eff}} \ket{\Psi} = E \ket{\Psi}$ can be decomposed into block matrix form:
\begin{equation}
\begin{pmatrix}
\hat{H}_{PP} & \hat{H}_{PQ} \\
\hat{H}_{QP} & \hat{H}_{QQ}
\end{pmatrix}
\begin{pmatrix}
\ket{\Psi_P} \\
\ket{\Psi_Q}
\end{pmatrix}
= E
\begin{pmatrix}
\ket{\Psi_P} \\
\ket{\Psi_Q}
\end{pmatrix},
\label{eq:block_matrix}
\end{equation}
where $\hat{H}_{PP} = \hat{P} \mathcal{M}_{\text{BdG}} \hat{P}$ represents the intrinsic physics, while $\hat{H}_{QQ} = \hat{Q} (\mathcal{M}_{\text{BdG}} + \Lambda) \hat{Q}$ dominates the high-energy sector due to the large penalty $\Lambda$. We emphasize that due to the complex Peierls phases in the Kitaev-Bose-Hubbard model, the Hamiltonian is generally non-Hermitian, implying $\hat{H}_{PQ} \neq \hat{H}_{QP}^\dagger$.

\subsection{2. Derivation of the Effective Hamiltonian}

Expanding Eq.~(\ref{eq:block_matrix}) yields two coupled equations:
\begin{align}
(E - \hat{H}_{PP}) \ket{\Psi_P} &= \hat{H}_{PQ} \ket{\Psi_Q}, \\
(E - \hat{H}_{QQ}) \ket{\Psi_Q} &= \hat{H}_{QP} \ket{\Psi_P}.
\end{align}
Assuming we are interested in the physical energy scale ($E \sim J, U \ll \Lambda$), the operator $(E - \hat{H}_{QQ})$ is invertible. We can express the spurious component as $\ket{\Psi_Q} = (E - \hat{H}_{QQ})^{-1} \hat{H}_{QP} \ket{\Psi_P}$. Substituting this back, we obtain an energy-dependent effective Hamiltonian for the physical sector:
\begin{equation}
\mathcal{H}_{\text{phys}}(E) = \hat{H}_{PP} + \hat{H}_{PQ} \frac{1}{E - \hat{H}_{QQ}} \hat{H}_{QP}.
\end{equation}
Since the penalty $\Lambda$ is the dominant energy scale, the inverse operator (Green's function) can be expanded perturbatively:
\begin{equation}
\frac{1}{E - \hat{H}_{QQ}} \approx \frac{1}{E - (E_{\text{spur}} + \Lambda)} \approx -\frac{1}{\Lambda} + \mathcal{O}\left(\frac{1}{\Lambda^2}\right).
\end{equation}
Substituting this approximation, the effective physical Hamiltonian becomes:
\begin{equation}
\mathcal{H}_{\text{phys}} \approx \hat{H}_{PP} - \frac{\hat{H}_{PQ} \hat{H}_{QP}}{\Lambda}.
\end{equation}
This result reveals that the contamination from the spurious sector introduces an energy shift to the physical eigenvalues:
\begin{equation}
\delta E_{\text{phys}} \approx - \frac{\langle \Psi_P | \hat{H}_{PQ} \hat{H}_{QP} | \Psi_P \rangle}{\Lambda}.
\end{equation}
Crucially, this shift scales as $1/\Lambda$. Even in our non-Hermitian system, the magnitude of this perturbative correction is strictly controlled by $\Lambda$. Therefore, by choosing $\Lambda \gg \max(J, U)$, the spurious modes are pushed to high energies while the physical spectrum remains asymptotically exact, validating the stability analysis presented in the main text.

\section{C. Geometric Stability of the Bubble Phase: Suppression of Percolation}

The stability of the Bubble Phase relies on a delicate balance between the bubble density and the bubble size. To stabilize a globally insulating yet compressible state, the system must maintain spontaneous spatial fragmentation. This requires that the superfluid islands (bubbles) remain sufficiently separated by insulating domain walls to prevent global percolation. In this section, we provide a detailed visualization of how increasing $\alpha$ or $\Delta$ drives the system towards an unstable superfluid regime by destroying this separation.

\begin{figure}[h]
    \centering
    \includegraphics[width=\linewidth]{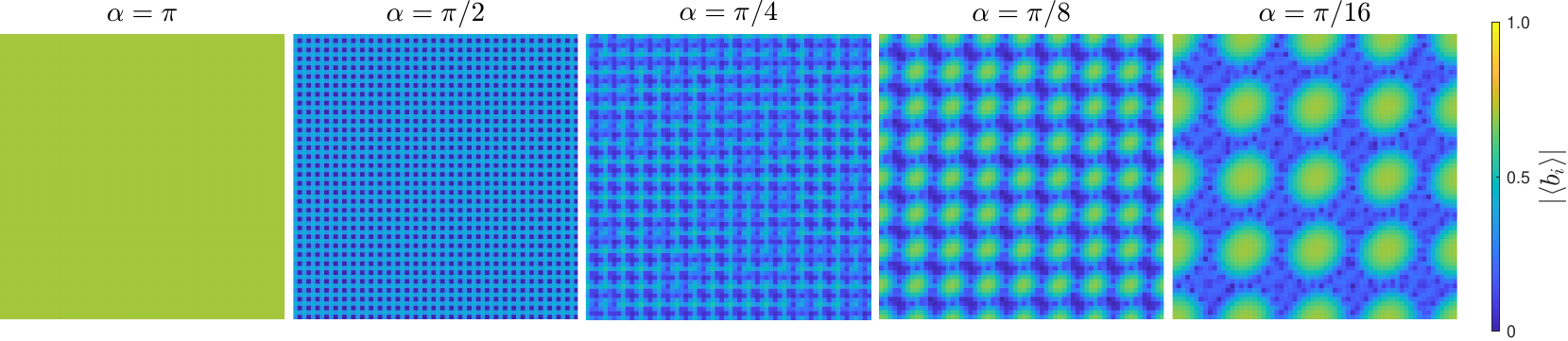} 
    \vspace{-0.1cm} \textsf{(a) Dependence on Magnetic Flux $\alpha$} \vspace{0.3cm}

    \includegraphics[width=\linewidth]{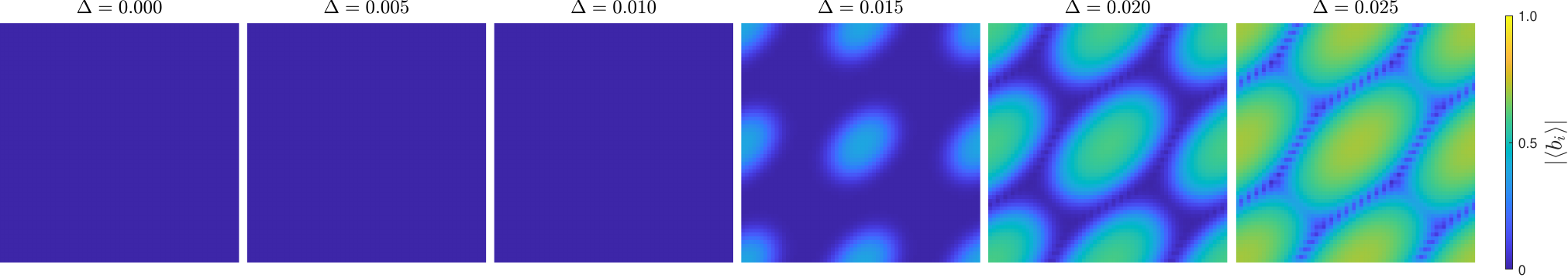}
    \vspace{-0.1cm} \textsf{(b) Dependence on Pairing Amplitude $\Delta$}
    
    \caption{Evolution of the real-space superfluid order parameter $|\langle b_i \rangle|$ demonstrating the percolation mechanism. 
    (a) Flux dependence (fixed $\Delta=0.005, J/U=0.025,\mu/U=0.8$): As the magnetic flux $\alpha$ increases from $\pi/16$ to $\pi$, the density of induced bubbles increases as $N_b \propto \alpha$. At high flux (e.g., $\alpha \ge \pi/8$), the bubbles become densely packed, reducing the mean separation and forcing a breakdown of the insulating barriers.
    (b) Pairing dependence (fixed $\alpha=\pi/16, J/U=0.025,\mu/U=0.8$): As the pairing amplitude $\Delta$ increases from $0$ to $0.025$, the spatial extent (radius) of each bubble expands. Even at a sparse flux density, a large $\Delta$ (e.g., $\Delta > 0.025$) causes neighboring bubbles to expand and overlap, initiating superfluid percolation.
    Therefore, the parameter choice used in the main text ($\alpha=\pi/16, \Delta=0.005$) is critical to maintain sparse, compact bubbles that prevent percolation.}
    \label{fig:supp_percolation}
\end{figure}

\subsection{1. Topological Crowding via Flux $\alpha$}
The number of superfluid bubbles is topologically determined by the number of magnetic flux quanta piercing the lattice, scaling as $N_b \approx (L \alpha/\pi)^2$. Figure~\ref{fig:supp_percolation}(a) illustrates this "crowding effect." 
\begin{itemize}
    \item At $\alpha=\pi/16$, the bubbles are sparse, separated by thick insulating domain walls, supporting a robust glassy phase.
    \item As $\alpha$ increases , the unit cell shrinks and the bubbles are forced into close proximity. The insulating regions are squeezed into thin filaments, drastically increasing the probability of tunneling between islands and leading to an unstable superfluid state.
\end{itemize}

\subsection{2. Bubble Expansion via Pairing $\Delta$}
While $\alpha$ sets the center-to-center distance, the pairing amplitude $\Delta$ (along with hopping $J$) determines the effective radius $R_b$ of each bubble. Figure~\ref{fig:supp_percolation}(b) demonstrates this "expansion effect" at a fixed low flux ($\alpha=\pi/16$).
\begin{itemize}
    \item At $\Delta=0$, the system is in a Mott insulating state.
    \item At $\Delta=0.005$ (our chosen value), clear, distinct bubbles form with finite size, but they remain strictly isolated.
    \item Increasing $\Delta$ promotes delocalization, causing the superfluid islands to expand spatially. This expansion progressively thins the insulating domain walls separating the bubbles. Once these barriers are effectively bridged, the localized clusters link up to form a global superfluid.
\end{itemize}

To realize the Intrinsic Bose Glass phase, it is insufficient to merely introduce frustration. One must simultaneously ensure a sparse density (low $\alpha$) and compact size (low $\Delta$) of the superfluid droplets. Our parameters $\alpha=\pi/16$ and $\Delta=0.005$ are specifically tuned to satisfy this condition, ensuring that transport is blocked by spontaneous spatial fragmentation.
\end{document}